\documentclass[conference]{IEEEtran}

\ifCLASSINFOpdf
   \usepackage[pdftex]{graphicx}
\else
\fi

\usepackage[tight,footnotesize]{subfigure}

\usepackage{stfloats}

\begin{document}

\title{Rectification by doped Mott-insulator junctions}

\author{\IEEEauthorblockN{Florian C. Sabou}
\IEEEauthorblockA{Department of Physics\\
Brown University\\
Providence, Rhode Island 02912-1843\\}
\and
\IEEEauthorblockN{Natalie Bodington}
\IEEEauthorblockA{Department of Physics\\
Brown University\\
Providence, Rhode Island 02912-1843\\}
\and
\IEEEauthorblockN{J. B. Marston}
\IEEEauthorblockA{Department of Physics\\
Brown University\\
Providence, Rhode Island 02912-1843\\}}

\maketitle

\begin{abstract}
Junctions of doped Mott insulators offer a route to rectification at frequencies beyond the terahertz range. Mott insulators have strong electronic correlations and therefore short timescales for electron-electron scattering.  It is this short time scale that allows for the possibility of rectification at frequencies higher than those of semiconductor devices that are limited by the slow diffusion of charge carriers.   We model a junction by a one dimensional chain of electrons with p- and n-doping on the two halves of the chain.  Two types of systems are investigated: spin polarized electrons with nearest-neighbor interaction, and spin-half electrons that interact via on-site  repulsion (the Hubbard model).  For short chains the many-body Schr\"odinger equation can be integrated numerically exactly, and when driven by an oscillating electromagnetic field such idealized junctions rectify, showing a preferred direction for charge transfer.  Longer chains are studied by the time-dependent density-matrix renormalization-group method, and also shown to rectify.  
\end{abstract}

\IEEEpeerreviewmaketitle

\section{Introduction}
\begin{figure*}
\centerline{\subfigure[]{ \includegraphics[width = 2in]{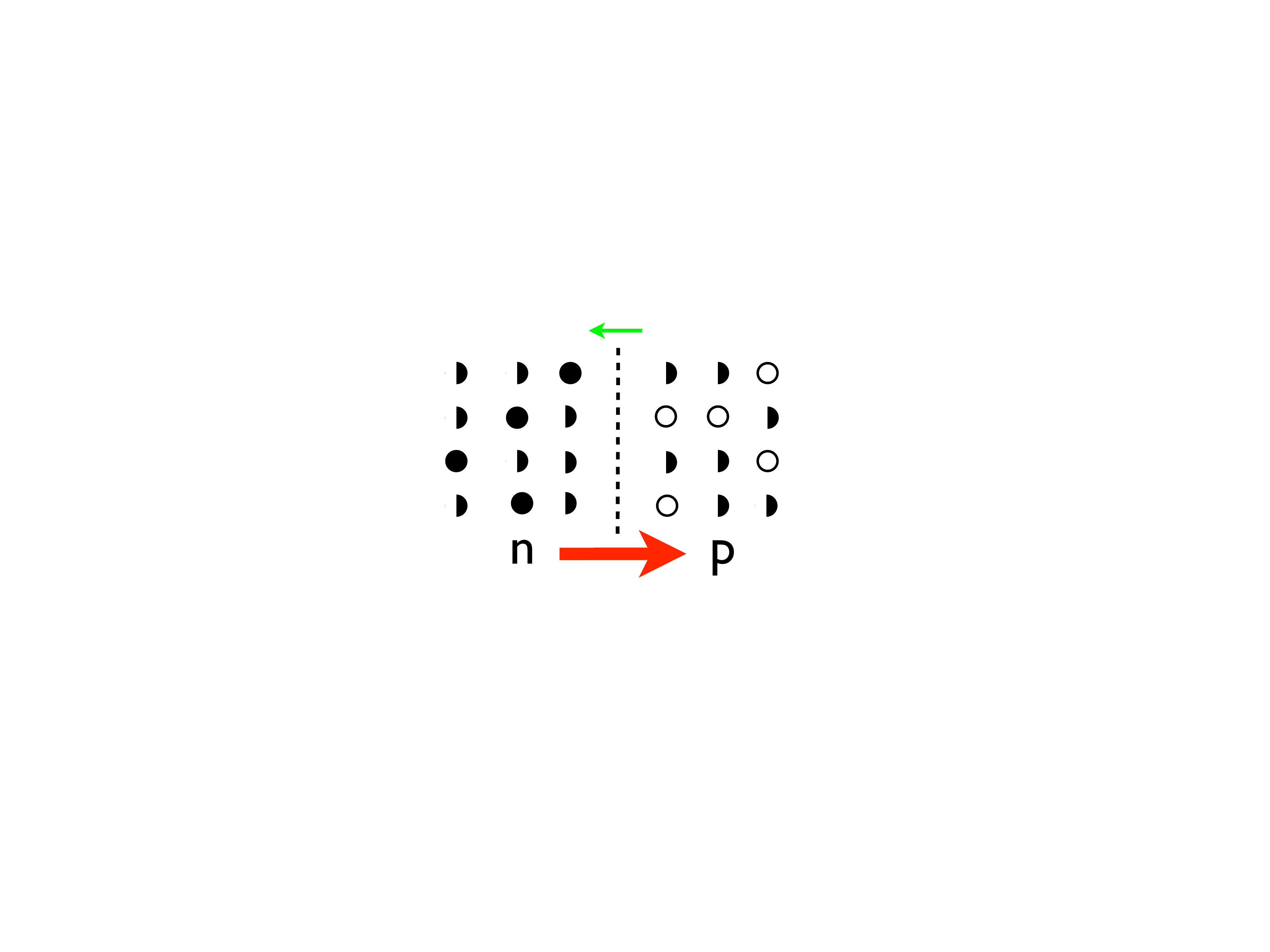}}
\subfigure []{ \includegraphics[width = 4in]{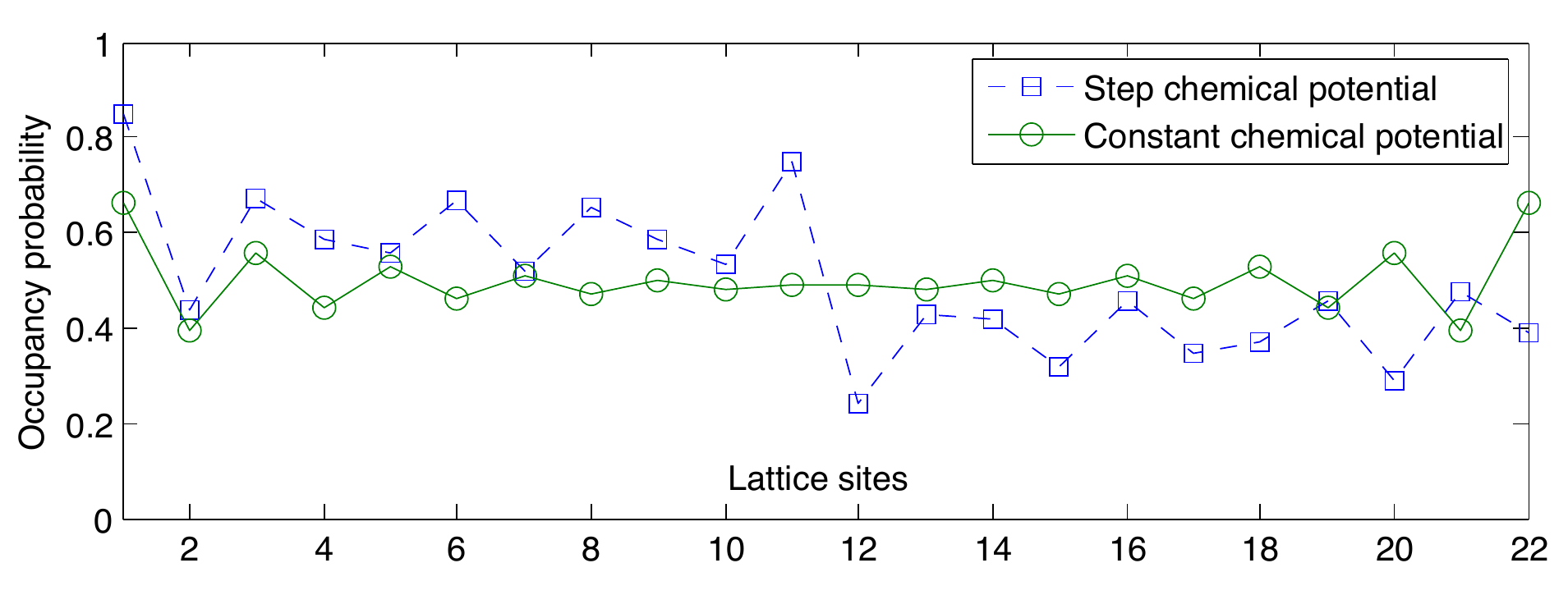}}}
\caption{(a) Schematic snapshot of a Mott insulator junction with atomic orbitals holding 0, 1, or 2 electrons (circles).  Electrons are driven preferentially to the right by an oscillating electric field acting in concert with strong intrasite Coulomb repulsion. (b) Ground state occupancy of the spin-polarized t-V chain, showing the n- (left) and p- (right) doped sides.  Here $V = 1$ and the local chemical potential $\mu = \pm 1$.}
\label{figure1}
\end{figure*}

Materials with strong electronic correlations offer tantalizing prospects for the construction of new
types of electronic devices \cite{braunecker2005}. One possibility is to replace the semiconducting host material of a pn
junction diode, or a semiconductor-metal Schottky diode, with doped Mott insulators \cite{orenstein2010} (see Fig. \ref{figure1}(a)). Such
a device could rectify at frequencies much higher than semiconducting diodes. The limiting factor in semiconducting diodes is the slow diffusion of charge carriers across the junction; 
by contrast strong correlations in Mott insulators can result in fast dynamics \cite{manousakis}. 

Transition metal oxides are often predicted to be conductors by band theory, but found to be
Mott insulators. Such oxides can show interesting behavior when driven by high frequency fields.  
For instance, an insulator-metal transition in vanadium dioxide induced by a terahertz electric field has recently been reported \cite{liu2012}.  Ambipolar field-effects
have been seen in metal-insulator-semiconductor transistors built from quasi-one-dimensional Mott
insulators \cite{hasegawa}. A pn junction consisting of a doped Mott-like insulator and a doped band
insulator (doped manganite and doped strontium titanate) exhibits rectification \cite{sun}. Heterojunctions of Mott insulators-band insulators show photovoltaic effects \cite{sun,qiu,luo}. There has also been theoretical progress in understanding junctions between Mott insulators \cite{manousakis,konemitsu1,konemitsu2}.

In this paper we show that idealized junctions between oppositely doped Mott insulators are able to rectify at very high frequencies. 
Numerically integrating the many-body Schr\"odinger equation for a one-dimensional model of spin-polarized fermions driven by an oscillating electric field 
shows that energy rises steadily as there is no mechanism for dissipation.  For
short chains we include dissipation in a phenomenological way by rotating the time axis into the complex plane.    Long Hubbard chains are simulated instead by means of the adaptive time-dependent density-matrix renormalization-group (tDMRG) method.    As shown below, sufficiently long chains can be simulated to detect rectification of a rapidly oscillating electric field before the junction current becomes contaminated by reflection off the open boundaries at the chain ends. 

In Section \ref{spinpolarized}  we present the spin-polarized one dimensional t-V model of the doped Mott insulator junction.  The necessity of two-body interactions, and hence correlations, for rectification is demonstrated.  Section \ref{rectification} details the response of the system to an oscillating electric field and the effects of dissipation and driving frequency on the non-equilibrium dynamics.  In Section \ref{tdmrg} the  tDMRG algorithm is used to simulate Hubbard chains.  Section \ref{conclusion} presents some conclusions.

\section{Spin-polarized model for junctions of doped Mott insulators}
\label{spinpolarized}
We first consider an idealized junction of doped Mott insulators that are modeled as one-dimensional chains of interacting spin-polarized fermions. The lattice sites of the chain can be occupied or empty, and there is a nearest-neighbor Coulomb repulsion between two adjacent occupied sites.  A computational advantage of spin-polarized electrons is that the Hilbert space of a chain of length $L$ is of dimension $2^L$, much smaller than the $4^L$ sized Hilbert space of unpolarized electrons.  Consequently chains of lengths exceeding 20 sites can be studied in a numerically exact manner.  At half filling, and in the atomic limit of small hopping amplitude, the  lowest energy state has the electrons occupying every other site in the chain in order to minimize the electrostatic repulsion.  The chain is insulating because the hopping of an electron to a neighboring site is inhibited by the nearest-neighbor electron-electron repulsion.  Doping away from half filling, by adding or removing electrons, results in a conducting state as now the electrons can hop without changing the total number of pairs of adjacent occupied sites.

When driven by an electric field of magnitude $\mathcal{E}$, the junction is described by the time-dependent t-V Hamiltonian:
\begin{eqnarray}\nonumber
H(t) &=& -\sum_{j} \left( c_{j}^{\dagger}c_{j+1} + H. c.\right ) + V \sum_{j} n_j n_{j+1}
\nonumber \\
 &+& \sum_{j} \mu_j(n_j-1/2) + \mathcal{E} \sin(\omega t) \sum_{j} j~ n_j
\label{hamiltonian}
\end{eqnarray}
where $c_j$ destroys a fermion at site $j$, and $n_j \equiv c_{j}^{\dagger}c_{j}$ is the number operator at site $j$.  The hopping amplitude has been set equal to unity ($t = 1$) so all other quantities, including time, are measured in units of $t$.  Doping is controlled by a local chemical potential $\mu_j$, and $V$ is the repulsive interaction between nearest-neighbor fermions.
Doping away from half-filling in opposing directions on the two halves of the 1D chain leads to electron-rich and electron-depleted halves as shown in Fig. \ref{figure1}(b).  

\section{Rectification of oscillating electric fields}
\label{rectification}
Rectification by the junction can be tested by driving it with an oscillating electric field $\mathcal{E}$  directed along the chain.  The many-body Schr\"{o}dinger's equation is integrated forward in time numerically using the Runge-Kutta 4th-order accurate algorithm, with the ground state (in the absence of the field) as the initial condition.  Deviations from unitary evolution are monitored by calculating the normalization of the many-body wave function $| \Psi(t) \rangle$, and are found to be negligible for the small time step that is used.

\subsection{Non-interacting limit}
\begin{figure*}
\centerline{\subfigure[]{ \includegraphics[width=3.0in]{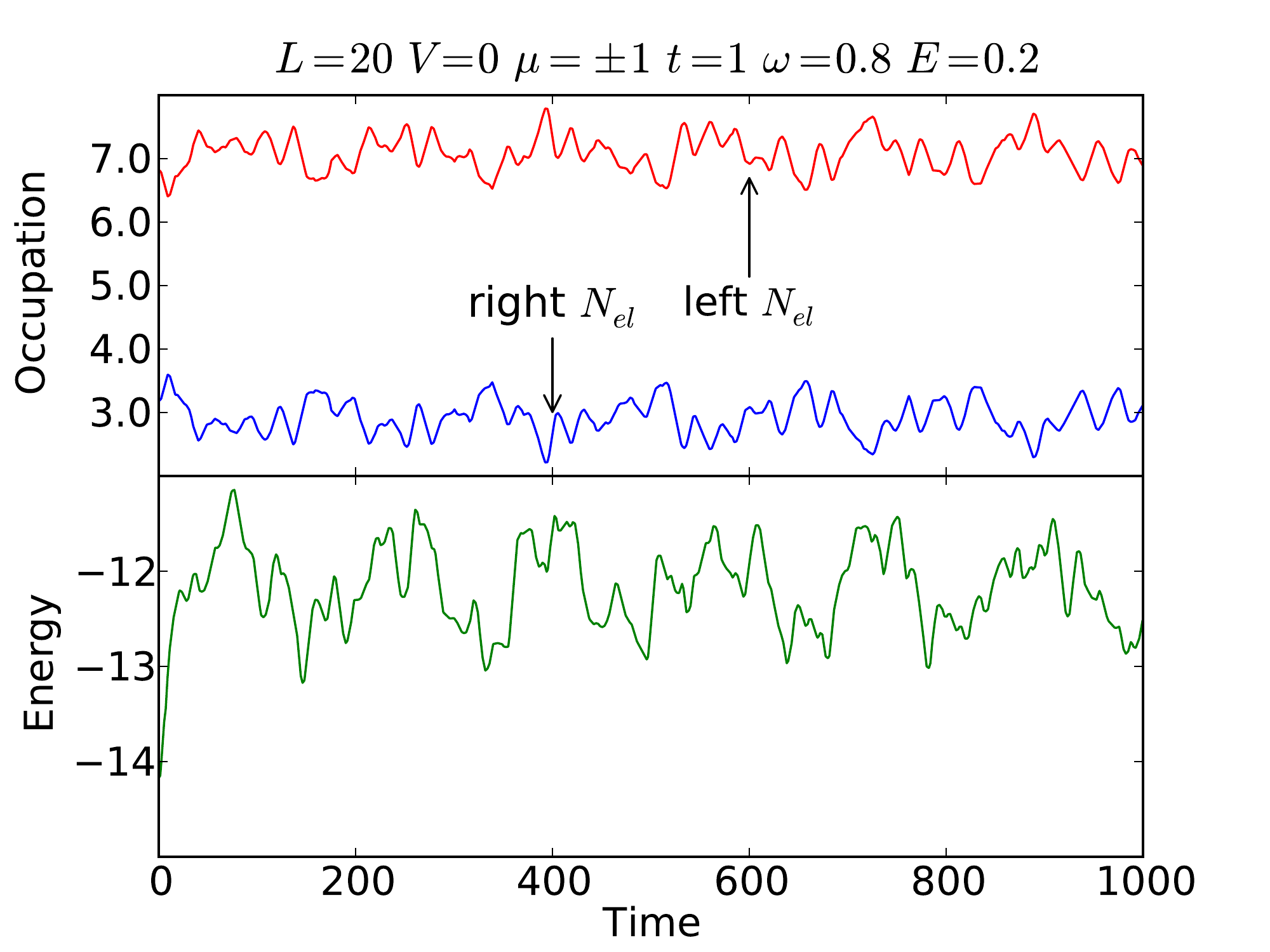}} 
\subfigure[] { \includegraphics[width=3.in]{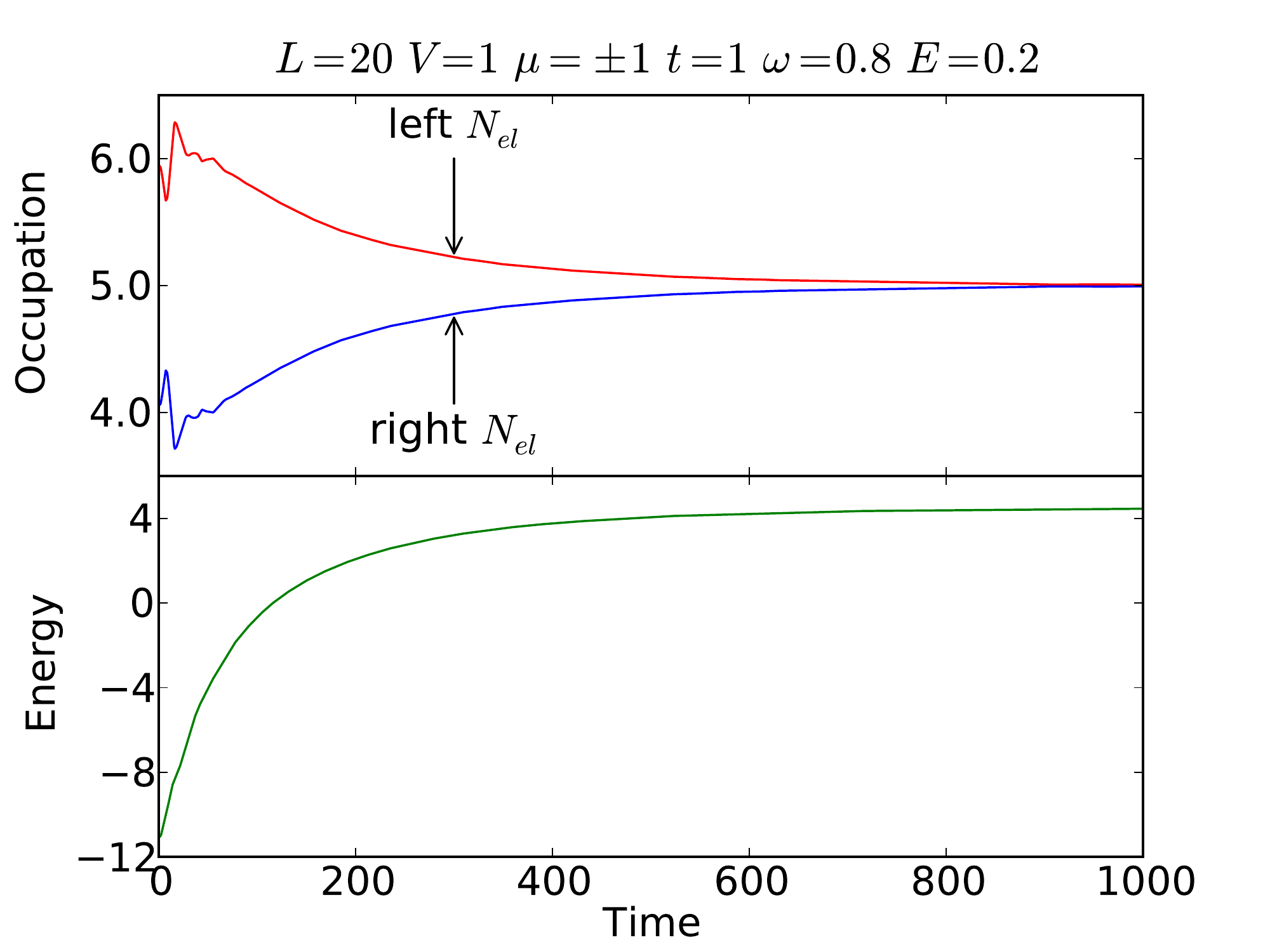}}}
\caption{Time evolution of electron occupancy (top) on the n and p sides of the
junction, and the energy $E(t)$ (bottom) for the case of an electric field oscillating at angular frequency $\omega = 0.8$.  Quantities shown in each plot are time-averaged over the period of the driving electric field.  (a) Non-interacting limit $V = 0$.  (b) Interacting electrons with nearest-neighbor repulsion $V = 1$ that shows net charge transfer (rectification).}
\label{figure2}
\end{figure*}

Rectification requires reflection asymmetry, but that in itself is not sufficient. To see this, consider first  the non-interacting limit $V=0$.  The response of a 20-site chain to the electric field is shown in Fig. \ref{figure2}(a).  The top panel shows that the occupancies on either side of the junction oscillate but no net-charge transfer occurs through the junction. The energy $E(t) \equiv \langle \Psi (t)| H(0) |\Psi(t)\rangle$ continues to fluctuate (bottom panel).  Reflection asymmetry by itself is insufficient for rectification to occur.  

\begin{figure*}
\centerline{\subfigure[]{ \includegraphics[width=3.0in]{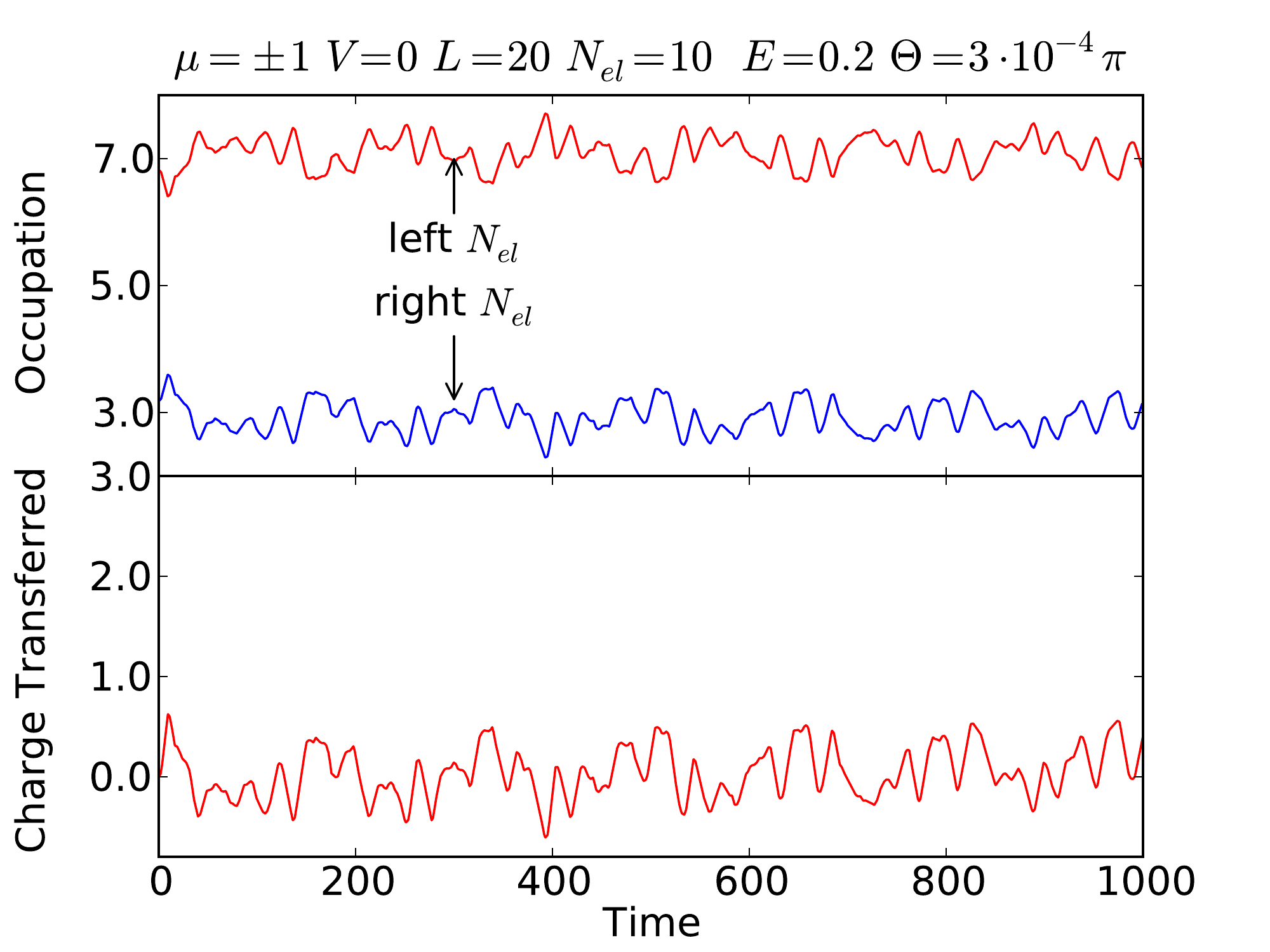}}  
\subfigure[] { \includegraphics[width=3.0in]{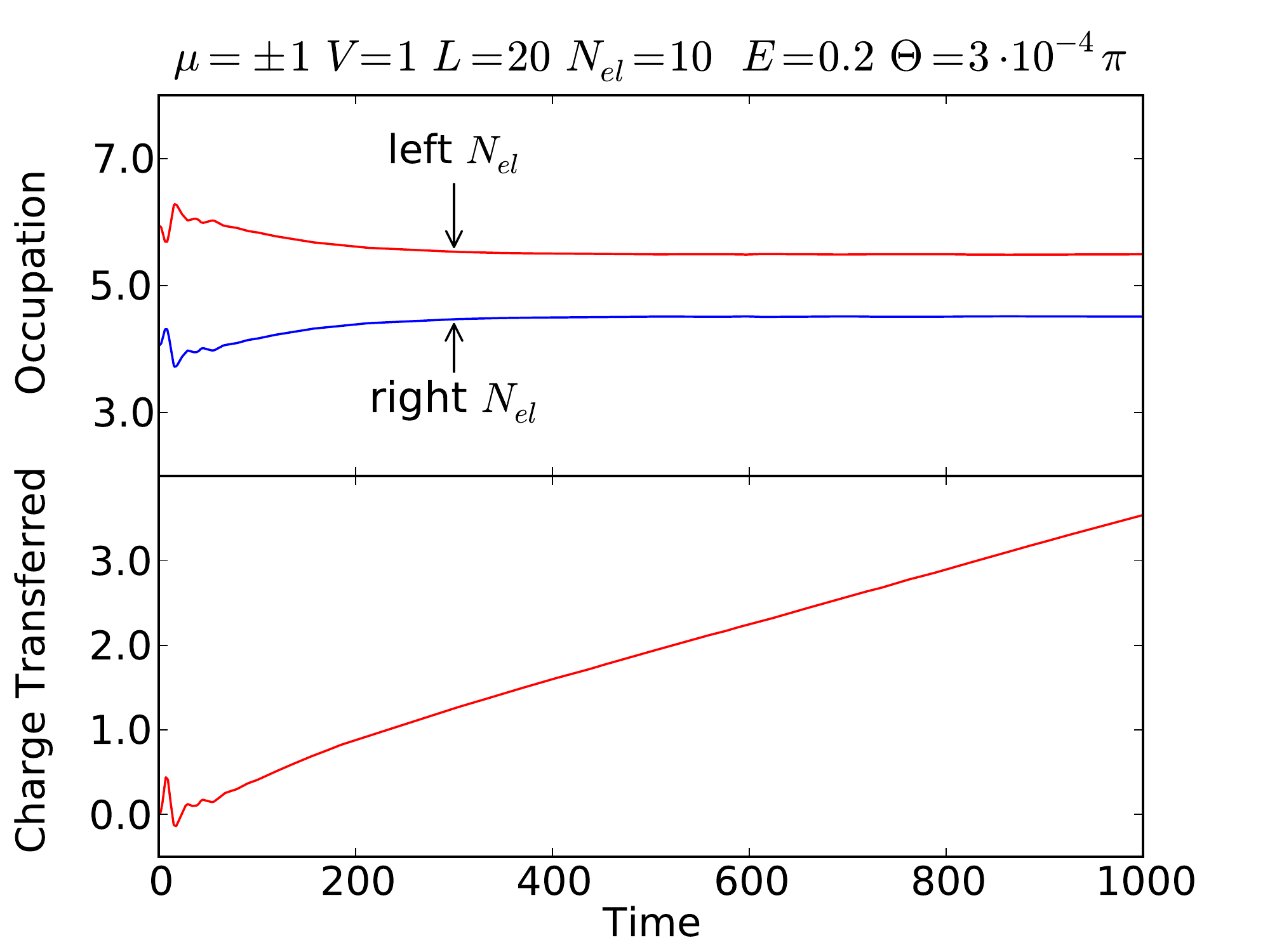}}}
\caption{Same as Fig. \ref{figure2} but with phenomenological dissipation modeled by $\theta = 3 \times 10^{-4} \pi$.   Note that rectification seen in (b) ceases when the
two-body interaction is turned off (a).}
\label{figure3}
\end{figure*}

\subsection{Interacting electrons}
Turning on the electron-electron interaction leads to a qualitative change.  The fermions are now strongly correlated and respond to the driving electric field quite differently.  Fig. \ref{figure2}(b) shows that the electron occupancy and energy $E(t)$ change smoothly after initial transients. Now  there is a net charge transfer through the junction, and the system steadily gains energy.   Eventually no more energy can be absorbed, and electrons are equally likely to be found in either side of the chain, reflecting their high energy.   
The time evolution is much smoother than in the noninteracting limit because the energy eigenvalues of $H(0)$ are spread out, and not concentrated at isolated values as they are at $V=0$.

\subsection{Dissipation}
\begin{figure}
\begin{center}
\includegraphics[width=3.5in]{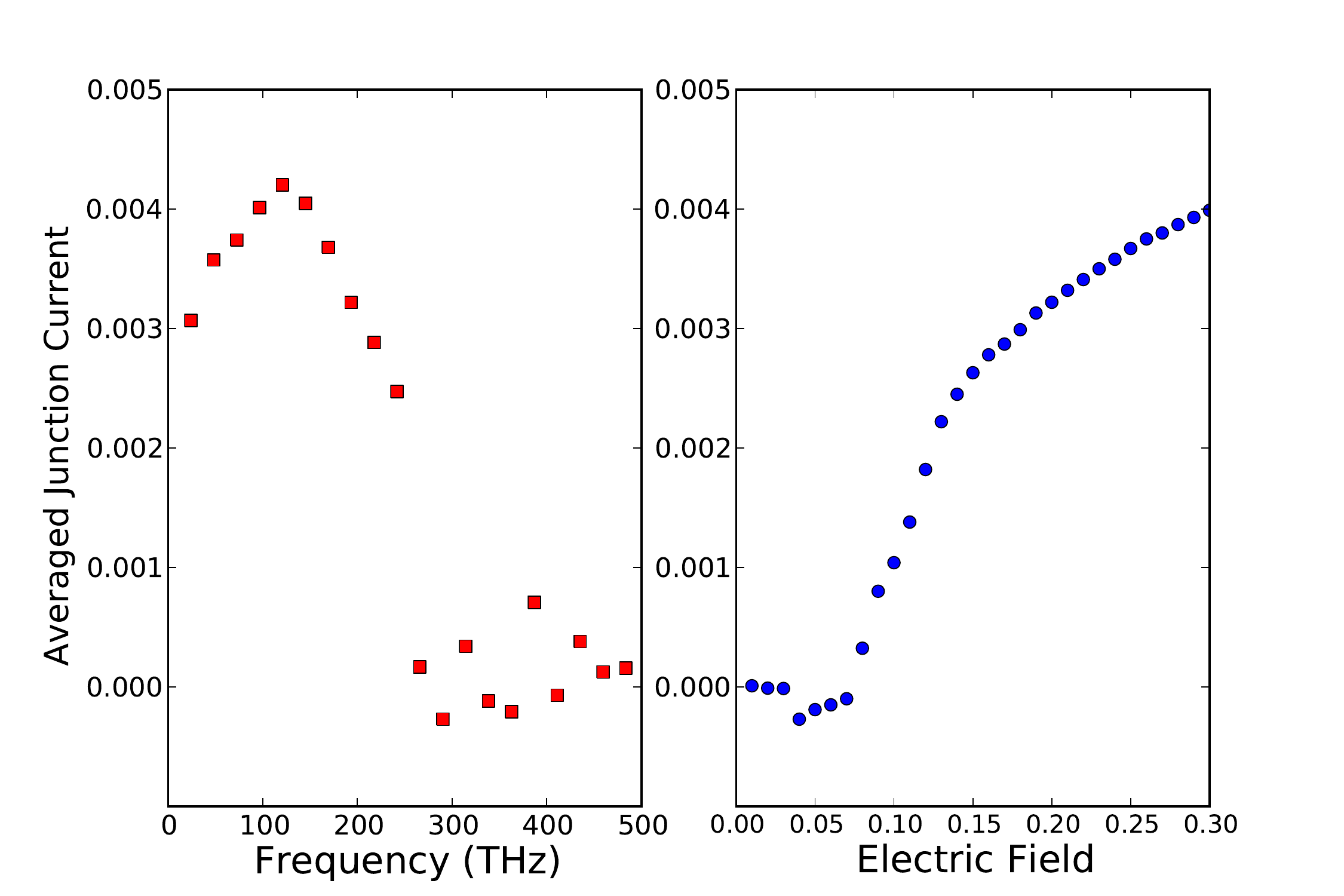}
\caption{Left: The cutoff in the frequency response is set by the electronic energy scales, illustrated here by $t = V = 2$ eV. Right: The dependence of the current on the magnitude of the electric field for the case of angular driving frequency $\omega = 0.8$.}
\label{figure4}
\end{center}
\end{figure}

As it stands, the energy rises steadily as there is no mechanism for dissipation in Eq. \ref{hamiltonian}.  Dissipation may be modeled phenomenologically by rotating the time axis into the complex plane. Unitary quantum evolution is then replaced by the modified dynamics:
\begin{equation}
i\hbar \frac{d}{dt} | \Psi(t)\rangle =  e^{-i\theta} H(t) | \Psi(t)\rangle 
\label{srt}
\end{equation}
where the state vector is renormalized to unity, $\langle\Psi(t)| \Psi(t)\rangle = 1$, periodically. For $\theta = \pi /2$ the integration is strictly along the imaginary axis and the system rapidly relaxes to the ground state.  Partial rotation away from the real time axis therefore acts as a restoring mechanism for the driven system.  Energy is steadily drawn out of the system, and a steady-state balance with the input power can be reached.   In contrast to unitary evolution of a finite system, here the current through the junction,
 \begin{equation}
J = i~ \left  \langle \Psi(t) \left |     \left(  c_{L/2}^{\dagger}c_{L/2-1} ^{}- c_{L/2-1}^{\dagger}c_{L/2}^{} \right)   \right | \Psi(t)\right \rangle, 
 \end{equation}
can reach a steady non-zero value, reflecting the continual renormalization of the wavefunction that effectively closes the circuit.  Fig. \ref{figure3} shows plots that correspond to those in Fig. \ref{figure2} but now with the added dissipation.  There is no rectification in the non-interacting $V = 0$ limit, but a DC current is evident for $V = 1$.  The frequency response of the rectification can be simulated by driving the system at different frequencies.  For qualitatively realistic values of the electronic parameters, $t$ = $V$ = 2 $eV$, the cuttoff frequency is of order $100$ THz as shown by the left panel of Fig. \ref{figure4}.  Likewise the $I-V$ response can be gauged by varying the magnitude of the driving electric field as shown in the right panel of Fig. \ref{figure4}. 

\section{TDMRG investigation of Hubbard chains}
\label{tdmrg}
The phenomenological model for dissipation described above is unsatisfactory because it lacks a basis in fundamental laws of quantum mechanics.  Better would be to simulate long chains by the time-dependent density-matrix renormalization-group (tDMRG) method.  More realistic models of doped Mott insulators, such as Hubbard chains and ladders, can be investigated by using tDMRG.   The nearest-neighbor interaction between spin-polarized electrons is replaced by on-site Coulomb repulsion between spin up and down electrons.  We use open-source code developed by Alvarez et al. \cite{alvarez} to study the flow of current through a junction of two idealized doped Mott insulators.    

\begin{figure}                                                                                                                                         
\begin{center}$                                                                                                                                           
\begin{array}{cc}                                                                                                                                          
\includegraphics[width=3.5in]{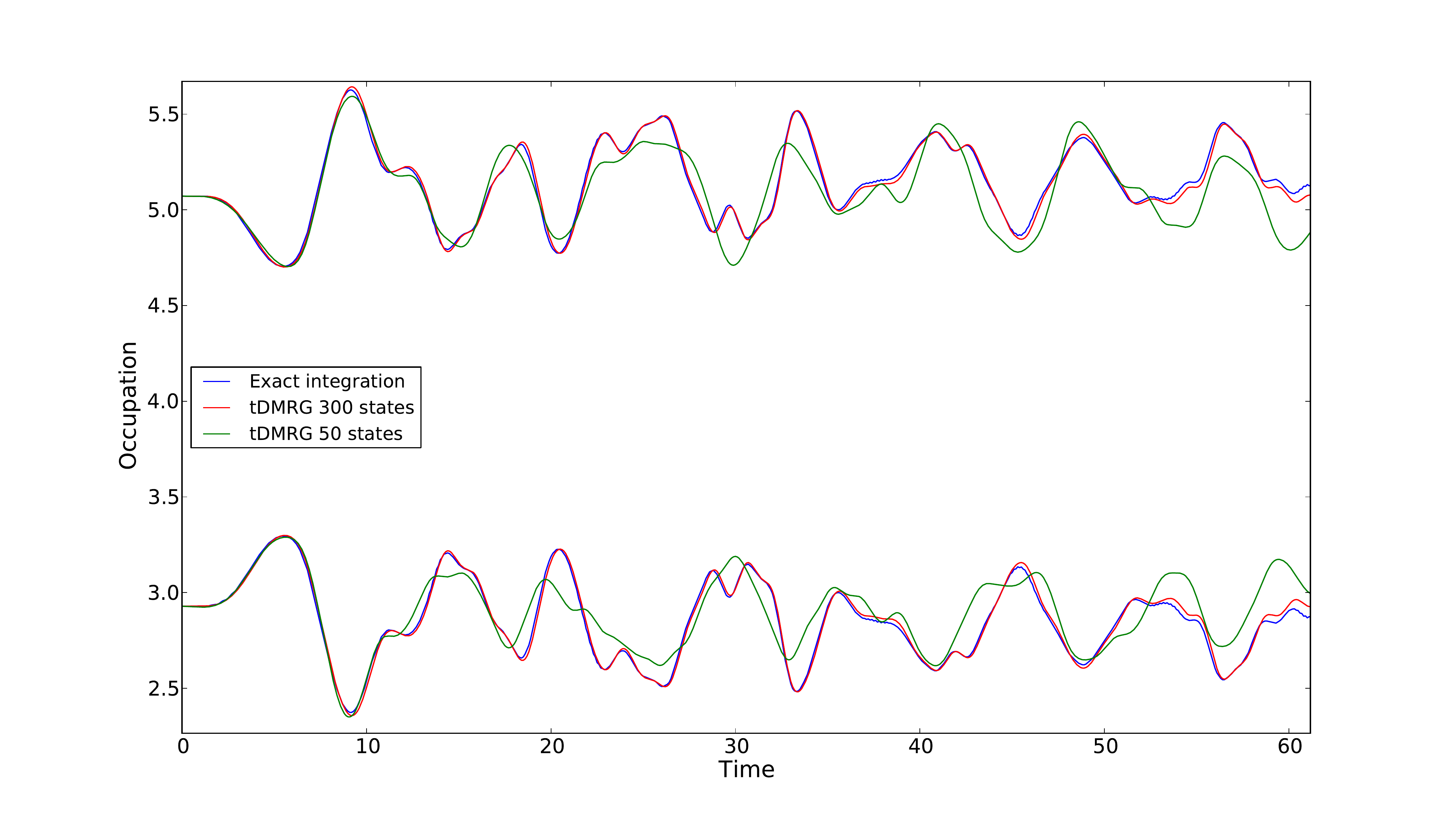}                                                                                     
\end{array}$                                                                                                                                               
\end{center}                                                                                                                                               
\caption{Comparison between the  tDMRG and numerically exact integration for a 10-site Hubbard chain. The system is driven by an oscillating electric field with angular frequency $\omega = 0.8$ and amplitude $\mathcal{E} = 0.1$.  The tDMRG blocks are of dimension $300$ and $50$, and the tDMRG time step is $0.1$.  The agreement with the exact solution deteriorates with time, especially in the case of small blocks which result in a truncation of the Hilbert space.}                                                                              
\label{figure5}                                                                                                                                                 
\end{figure}

\begin{figure}
\begin{center}
\includegraphics[width=3in]{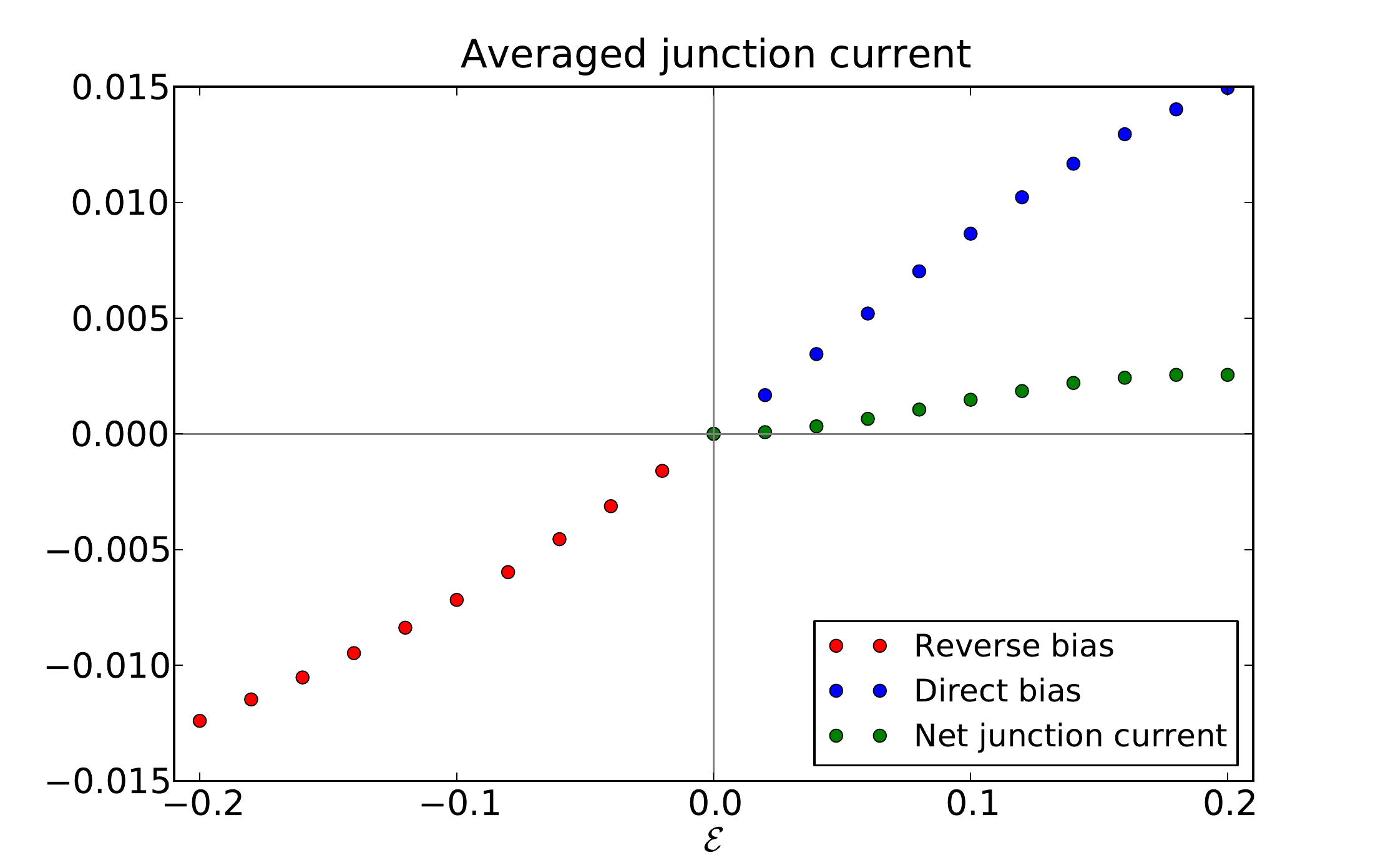}
\caption{tDMRG simulation (with block size of $300$) of a 40-site Hubbard chain with $\mu = \pm 0.5$, Coulomb interaction $U = 2$, and driving frequency $\omega = 0.8$.  The driving electric field $\mathcal{E}(t)$ grows sinusoidally from zero to a maximum,  and an averaged junction current is calculated over this time interval.  The response of the junction to this field, and one of opposite polarity, is shown.  The net current quantifies the amount of rectification.}
\label{figure6}
\end{center}
\end{figure}

DMRG has been shown to be an  accurate method to find ground state properties, but the time-dependent generalization may be seen to diverge from the exact solution due to the truncation of the Hilbert space and the non-infinitesimal time step.  In Fig. \ref{figure5} we compare tDMRG with two different block sizes against numerically exact time integration of a short 10-site Hubbard chain. Agreement is excellent at short times but the occupancies diverge at later times.  For the blocks of dimension $300$ the discrepancy is small and due solely to the time step; small blocks of dimension $50$ introduce further errors due to the truncation of the Hilbert space.  

tDMRG with blocks of dimension $300$ is used in a preliminary investigation of the response of a 40-site junction immediately after the oscillating electric field begins to turn on.  The junction current is calculated for electric fields of opposite initial polarity; differences in the the magnitude of the current then provide a measure of rectification.  Fig. \ref{figure6} shows the forward, reverse, and net junction current as a function of the amplitude of the driving electric field. The finite value of the net current shows that the model junction responds asymmetrically to changes in the polarity of initial applied field.   Whether or not such rectification is sustained over longer times is a question that remains to be addressed.

\section{Conclusion}
\label{conclusion}
Idealized models of junctions between doped Mott insulators have been shown to exhibit rectification up to frequencies of order the electronic scale. Short chains of spin-polarized electrons with and without phenomenological dissipation were investigated by numerically exact time integration.  Rectification ceases in the non-interacting limit, as expected, because it requires not only broken reflection symmetry but also strong electronic correlations.  Longer Hubbard model chains were simulated using the tDMRG algorithm without added dissipation. Further studies of longer chains and more realistic multi-channel models can test the potential for devices made from doped transition metal oxides such as VO$_2$, LaVO$_3$, and NdNiO$_3$ to rectify high-frequency electric fields.  New classes of applications may be possible with such devices.

\section*{Acknowledgment}
{\footnotesize{We thank Vladan Mlinar and Domenico Pacifici for helpful discussions, and Peter Weber for support.  This work was funded in part by DOE DE-SCOOO1556 and NSF DMR-0605619.  The open-source tDMRG code was developed by  Gonzalo Alvarez at the Center for Nanophase Materials Sciences, which is sponsored at Oak Ridge National Laboratory by the Scientific User Facilities Division, Office of Basic Energy Sciences, U.S. Department of Energy.}}

\end{document}